\input{aipcheck}

\documentclass[
    ,final            
  ]
  {aipproc}

\layoutstyle{8x11single}

\usepackage{color}

\begin{document}

\quad{PI/UAN-2013-560FT}

\title{The Vector Curvaton}


\author{Andr\'es A. Navarro}{
  address={Escuela de F\'{\i}sica, Universidad Industrial de Santander,\\
Ciudad Universitaria, Bucaramanga 680002, Colombia.}
}

\author{Yeinzon Rodr\'{\i}guez}{
  address={Escuela de F\'{\i}sica, Universidad Industrial de Santander,\\
Ciudad Universitaria, Bucaramanga 680002, Colombia.}
,altaddress={Centro de Investigaciones en Ciencias B\'asicas y Aplicadas, Universidad Antonio Nari\~no,\\
Cra 3 Este \# 47A - 15, Bogot\'a D.C. 110231, Colombia.}
}


\begin{abstract}
We analyze a massive vector field with a non-canonical kinetic term in the action, minimally coupled to gravity, where the mass and kinetic function of the vector field vary as functions of time during inflation. The vector field is introduced following the same idea of a scalar curvaton, which must not affect the inflationary dynamics since its energy density during inflation is negligible compared to the total energy density in the Universe. Using this hypothesis, the vector curvaton will be solely responsible for generating the primordial curvature perturbation $\zeta$. We have found that the spectra of the vector field perturbations are scale-invariant in superhorizon scales due to the suitable choice of the time dependence of the kinetic function and the effective mass during inflation. The preferred direction, generated by the vector field, makes the spectrum of $\zeta$ depend on the wavevector, i.e. there exists statistical anisotropy in $\zeta$. This is discussed principally in the case where the mass of the vector field increases with time during inflation, where it is possible to find a heavy field ($M\gg H$) at the end of inflation, making the particle production be practically isotropic; thus, the longitudinal and transverse spectra are nearly the same order which in turn causes that the statistical anisotropy generated by the vector field is within the observational bounds.

\end{abstract}

\maketitle


\section{Introduction}
The standard cosmological model known as the Big Bang accounts for the primordial abundance of light elements and the observed Cosmic Microwave Background (CMB) radiation seen for first time by Penzias and Wilson \cite{2}. However, despite of the predictions of the Big Bang, this theory has difficulties such as the flatness, horizon and unwanted relics problems \cite{3}. The cosmological inflationary model  proposed by A. Guth in 1981 \cite{4} is the most elegant solution to these problems. In addition to solving the above mentioned problems, the model is also able to generate a discriminant quantity among different inflationary models: the primordial curvature perturbation $\zeta$, which is responsible for the generation of the large scale structure  (galaxies and clusters of galaxies) \cite{3}. During inflation, the quantum fluctuations of the fields are amplified, making them classic and almost constant after the cosmological scales leave the horizon \cite{4a}. These fluctuations generate small inhomogeneities in the energy density of the fields during inflation, which in turn are related to $\zeta$ and the fractional difference of temperature in the CMB by the Sachs-Wolfe effect \cite{5}:
\begin{equation}
\left(\frac{\delta T}{T_{0}}\right)_{k}= -\frac{1}{2}\left(\frac{\delta\rho}{\rho_0}\right)_{k}=-\frac{1}{5}\zeta_{k}\approx 10^{-5},
\end{equation}
where the subindex ``$0$'' means the average of the respective quantity, and $k$ denotes the Fourier mode whose wavelength is of the size of the observable universe.

The traditional way to generate the primordial curvature perturbation $\zeta$  is by means of a scalar field; this way preserves the homogeneity and isotropy at large scales. The double task of this scalar field during inflation is: to solve the standard cosmological problems \cite{3} and to generate the large-scale structure; this is very restrictive, discarding inflationary models well motivated in particle physics that cannot generate the observed large-scale structure \cite{8}. An alternative way is to consider an additional scalar field called the curvaton \cite{9, 10, 11}, which only generates $\zeta$ after the inflationary period; therefore, during inflation the energy density associated to the curvaton must be negligible compared to the total energy density in the Universe; thus, the curvaton is not involved in the inflationary dynamics but generates the large-scale structure after inflation when the energy density can be increased and nearly dominate the total energy density in the Universe.\\

The inflationary models constructed from scalar fields have been a great tool to predict the characteristics in the CMB, but while the LHC confirms that the new discovered particle  corresponds to a scalar or vector field, uncertainty about the existence of fundamental scalar fields is present \cite{12,13}; in addition, the WMAP data suggested the existence of a preferred direction in the CMB maps \cite{15,16,17}. Inflationary models constructed from scalar fields cannot describe this new feature; therefore, models which include vector fields are more attractive to explain the possible preferred direction in the Universe. However, in the traditional and accepted cosmological models, the primordial curvature perturbation    $\zeta$ is adiabatic (fluids during the inflationary epoch must be non-interacting), practically gaussian (the distribution function governing the curvature perturbation corresponds to a Gaussian distribution function) and practically scale invariant (the spectrum of $\zeta$ should be independent on the wavenumber), as long as they preserve statistical homogeneity and statistical isotropy; this last property may be violated by introducing vector fields during inflation, making the spectrum of $\zeta$ not invariant under spatial rotations  \cite{18,19}. One possible way to parameterize this departure from statistical isotropy is by redefining the spectrum of $\zeta$ \cite{20}:
\begin{equation}\label{eq2}
{ P}_{\zeta}(\textbf{k})={ P}^{\textrm{iso}}_{\zeta}(k)\left[1+g_{\scriptscriptstyle\zeta}(\hat{\textbf{d}}\cdot\hat{\textbf{k}})^{2}\right],
\end{equation}
where ${ P}^{\textrm{iso}}_{\zeta}$ is the spectrum of the primordial curvature perturbation $\zeta$ which depends only on the magnitude of the wavevector $\textbf{k}$, $\hat{\textbf{k}}$ is a unit vector in the direction of $\textbf{k}$, $\hat{\textbf{d}}$ is the unit vector along the preferred direction, and $g_{\scriptscriptstyle\zeta}$ is the level of statistical anisotropy. Studies from the WMAP experiment have established a range for  $g_{\scriptscriptstyle\zeta}$: $g_{\scriptscriptstyle\zeta}=0.290\pm 0.031$ \cite{15} which excludes statistical isotropy at more than $9\sigma$; however, the preferred direction in the anisotropies in the CMB lies near the plane of the solar system, so the authors of Ref. \cite{15} suggest that this effect may be due to unresolved systematic errors (see e.g. Ref. \cite{26ab}).

With the above in mind, we consider a vector field during inflation with a non-canonical kinetic term in the Lagrangian which does not affect the inflationary dynamics, following the spirit of the scalar curvaton scenario \cite{9,10,11}. This is done by making the vector field subdominant during inflation and has the effect of not generating excessive statistical anisotropy  that would disagree with the observations.
\section{the setup}

We consider the vector curvaton scenario with a Lagrangian for the vector field during inflation of the form
\begin{equation}\label{3.28}
L=-\frac{1}{4}fF_{\mu\nu}F^{\mu\nu}+\frac{1}{2}m^{2}A_{\mu}A^{\mu},
\end{equation}
where $f=f(t)$ is the kinetic term,  $m=m(t)$ is the mass of vector field, both being functions of the cosmic time, and $F_{\mu\nu}=\partial_{\mu}A_{\nu}-\partial_{\nu}A_{\mu}$ is the field strength tensor. The above Lagrangian density can be of a massive Abelian gauge field, in which case $f$ is the gauge kinetic function; however, we need not restrict ourselves to gauge fields only. If no gauge symmetry is considered, the argument in support of the above Maxwell-type kinetic term is that it is one of the few (three) choices \cite{106} which avoids introducing instabilities, such as ghosts \cite{56,57,58}.
First, we are in a period of cosmic expansion in which we assume that the contribution to the total energy density of the Universe due to the vector field is small and can be neglected; therefore, the inflationary epoch is isotropic and is driven by the inflaton scalar field. It is also considered that inflation is de Sitter, i.e. $H\approx$ constant.

The vector field perturbations during inflation live on a spatially flat, homogeneous, and isotropic background; therefore, the metric that will be used is the Friedmann-Robertson-Walker (FRW) metric:
\begin{equation}\label{FRW2}
ds^{2}=-dt^{2}+a^{2}(t)\delta_{ij}dx^{i}dx^{j},
\end{equation}
where the spatial coordinate system is cartesian.
However, this assumption is not completely true because the presence of a vector field during inflation induces an anisotropic metric of the Bianchi I type, but, as the vector field is subdominant during inflation (the energy density of the vector field is negligible during inflation), the anisotropic expansion generated by the vector field is negligible, so a good approximation to the background metric is the FRW one. With all the above, the physical  and canonically normalized vector field is defined as
\begin{equation}\label{3.38}
\textbf{W}=\frac{\sqrt{f}\textbf{A}}{a}.
\end{equation}
Perturbing the equations of motion, obtained from the Lagrangian in Eq. (\ref{3.28}), as described in Refs. \cite{81,27,28,29}, we found that the evolution equations for the transverse and longitudinal modes of the vector field perturbations are respectively
\begin{eqnarray}\label{3.39}
\left\lbrace \partial_{t}^{2}+3H\partial_{t}+\frac{1}{2}\left[ \frac{1}{2}\left( \frac{\dot{f}}{f}\right) ^{2}-\frac{\ddot{f}}{f}-\frac{\dot{f}}{f}H+4H^{2}\right] \right.
+\frac{m^{2}}{f}+\left( \frac{k}{a}\right) ^{2}\bigg\}\delta{{\textbf W}}^{\perp}=0,
\end{eqnarray}

\begin{eqnarray}\label{3.40}
\left\lbrace  \partial_{t}^{2}+\left[ 3H +\left( 2H+2\frac{\dot{m}}{m}-\frac{\dot{f}}{f}\right)\frac{(\frac{k}{a})^{2}}{(\frac{k}{a})^{2}+\frac{m^{2}}{f}}\right]\partial_{t} \right. 
+\frac{1}{2}\left[ \frac{1}{2}\left(\frac{\dot{f}}{f}\right) ^{2}-\frac{\ddot{f}}{f}-\frac{\dot{f}}{f}H+4H^{2}\right]\nonumber\\
\left. +\left( H-\frac{1}{2}\frac{\dot{f}}{f}\right)\left( 2H+2\frac{\dot{m}}{m}-\frac{\dot{f}}{f}\right) \frac{(\frac{k}{a})^{2}}{(\frac{k}{a})^{2}+\frac{m^{2}}{f}}+\frac{m^{2}}{f}+\left( \frac{k}{a}\right) ^{2}\right\rbrace \delta{{\textbf W}}^{\parallel}=0,
\end{eqnarray}
where the dots mean derivatives with respect to the cosmic time.

\section{Particle Production Process}
To study the particle production process associated to the vector field during inflation, we first need to promote $\delta\textbf{W}$ to a quantum operator. After that, we expand $\delta{\hat \textbf{W}}$ in creation and annihilation operators as
\begin{equation}
\delta{\hat \textbf{W}}=\int\frac{d^{3}k}{(2\pi)^{3}}\sum_{\lambda}\big[{\textbf e}_{\lambda}\hat a_{\lambda}({\textbf k})w_{\lambda}(t,k)e^{i\textbf{k}\cdot\textbf{x}}+
{\textbf e}^{*}_{\lambda}\hat a_{\lambda}^{\dagger}({\textbf k})w_{\lambda}^{*}(t,k)e^{-i\textbf{k}\cdot\textbf{x}}
\big],
\end{equation}
where $\lambda$ denotes the left, right, and longitudinal polarizations, and ${\bf e}_\lambda$ corresponds to the polarization vectors. The most convenient choice is the circular polarization for which two transverse vectors have different chirality. Because both of them transform differently under rotations, the rotational invariance of the Lagrangian prevents any couplings among them. Choosing the coordinate $z$ axis to point along the direction of $\textbf{k}$, the circular polarization vectors ${\bf e}_\lambda$ take the form \cite{18}
\begin{equation}
e_{L}\equiv\frac{1}{\sqrt{2}}(1,i,0),\,\,\,\,\,e_{R}\equiv\frac{1}{\sqrt{2}}(1,-i,0),\,\,\,\,\,e_{\parallel}\equiv(0,0,1).
\end{equation}
In addition, we impose canonical quantisation via
\begin{equation}
\left[ \hat{a}_{\lambda}(\textbf{k}),\hat{a}^{\dagger}_{\lambda '}(\textbf{k}') \right]=(2\pi)^{3}\delta(\textbf{k}-\textbf{k}')\delta_{\lambda\lambda '}.
\end{equation}
The equations given in (\ref{3.39}) and (\ref{3.40}) are linear and are satisfied by the corresponding modes $w_{\lambda}(t,\textbf{k})$. For an analysis of the spectra of perturbations, we need the solution of the field equations for modes with some appropriate initial conditions, so it is possible to obtain adequate restrictions over $f$ and $m$ in order that they provide a scale-invariant spectrum. The particle production process was studied in Ref. \cite{27} where it was found that the transverse spectra of perturbations will be scale-invariant when
\begin{equation}\label{11}
f\propto a^{-1\pm3}\quad\textrm{and}\quad M_{*}\ll H,
\end{equation}
where the subscript "$^{*}$" indicates the time when the scales leave the horizon and $M\equiv m/\sqrt{f}$ is the effective mass of the vector field $\textbf{W}$.

In the case of the longitudinal component, the spectrum will be scale-invariant when \cite{27}
\begin{equation}
m\propto a.
\end{equation}
Thus, the transverse and longitudinal spectra are scale invariant and take the following values
\begin{equation}
P_{+}=\bigg( \frac{H}{2\pi} \bigg)^{2}\quad\textrm{and}\quad P_{\parallel}=9\bigg(\frac{H}{M}\bigg)^{2}\bigg(\frac{H}{2\pi}\bigg)^{2},
\end{equation}
where $P_{+}$ and $P_{\parallel}$ correspond to the parity-conserving and longitudinal spectra respectively (defined as $P_\pm = (P_R \pm P_L)/2$).

\section{Evolution of the Zero Mode}
To find and analyze the evolution of the energy density and pressure associated to the vector field during inflation and after its end, it is necessary to find the energy-momentum tensor of the vector field. This is important to guarantee that the vector field does not generate anisotropic expansion. The energy-momentum tensor can be written as (assuming the preferred direction of the vector field points along the $z$ axis)
\begin{equation}
T_{\mu}^{\nu}=\textrm{diag}(\rho_{W},-p_{\perp},-p_{\perp}, p_{\perp}),
\end{equation}
where
\begin{equation}\label{3.81a}
\rho_{W}\equiv\rho_{\textrm{kin}}+V_{W},\quad p_{\perp}\equiv\rho_{\textrm{kin}}-V_{W},
\end{equation}
being
\begin{eqnarray}
\rho_{\textrm{kin}}&\equiv&-\frac{1}{4}fF_{\mu\nu}F^{\mu\nu}=\frac{1}{2}\frac{f\dot{A}^{2}}{a^{2}}=\frac{1}{2}[\dot{W}+\left(1-\frac{\alpha}{2}\right)HW]^{2},\label{3.81}\\
V_{W}&\equiv&-\frac{1}{2}m^{2}A_{\mu}A{\mu}=\frac{1}{2}\frac{m^2A^2}{a^2}=\frac{1}{2}M^2W^2,\label{3.82}
\end{eqnarray}
where $W\equiv|\textbf{W}|$.
From the above, it is important to find the evolution of the zero mode of the vector field $\textbf{W}$ to determine how the energy density and pressure evolve. From the Euler-Lagrange equations, the FRW metric, the definition of the canonically normalised vector field in Eq. (\ref{3.38}), the ansatz for $f$ of the form $f\propto a^{\alpha}$, and the expectation that inflation homogenises the vector field (making $\partial_{i} A_{\mu} = 0$ for $\mu = 0,1,2,3$), the evolution equation for the zero mode becomes
\begin{equation}\label{3.84}
\ddot{\textbf{W}}+3H\dot{\textbf{W}}+\left[ \left( 1-\frac{\alpha}{2}\right)\dot{H}-\frac{1}{4}H^{2}(\alpha +4)(\alpha -2)+M^2 \right] \textbf{W}=0,
\end{equation}
where $\alpha$ is a constant with values: $\alpha=-1\pm3$, so that  the transverse spectra are scale-invariant.

For the case $f\propto a^{-4}$, the  effective mass of the vector field scales as $M\propto a^{3}$; therefore, the solution for Eq. (\ref {3.84}) is
\begin{eqnarray}
W=W_{0}\left(\frac{a}{a_{0}} \right) ^{-3}\sqrt{2}\textrm{cos}\left(\frac{M}{3H}\pm\frac{\pi}{4} \right). 
\end{eqnarray}
Using the above result in Eqs. (\ref{3.81a}), (\ref{3.81}), and (\ref{3.82}), we found that the energy density of the vector field during inflation is
\begin{equation}
\rho_{W}=M_{0}^{2}W_{0}^{2},
\end{equation}
where $W_{0}$ is an initial condition for the vector field  ${\bf W}$ and $M_{0}$ is an initial condition for $M$, both of which assume the condition of initial equipartition of energy that corresponds to $(\rho_{\textrm{kin}})_{0}\simeq(V_{W})_{0}$, where the subscript '0' indicates the values at some initial time, e.g. near the onset of inflation.

In the case of $f\propto a^{2}$, the effective mass of the vector field is constant during inflation, so the solution to Eq. (\ref{3.84}) is
\begin{equation}
W=a_{0}^{-3}C\left[ \left(\frac{a_{0}}{a} \right)^{3}\pm \frac{3H}{M_{0}} \right]\simeq\,\textrm{contant}\simeq W_{0},
\end{equation}
because, after the onset of inflation, $(a_{0}/a)^{3}\ll 1\ll 3H/M_{0}$, $H$ being a constant and $C$ being a constant of integration. Thus, $W$ remains constant;  in addition, since $M$ is constant, $V_W$ also remains constant. However, since $\rho_{\rm kin} \propto a^{-6}$ (as shown in Ref. \cite{27}), and the initial equipartition of energy is also assumed, we find that, during inflation, $\rho_{\rm kin} \ll V_W$. Therefore, 
\begin{equation}
\rho_{W} \approx V_W \simeq M_{0}^{2}W_{0}^{2}.
\end{equation}
These results show that the density is independent on the mass of the vector field, which is valid for the regimes: $M\ll H$ and $M\gg H$. Consequently, if the density of the vector field is subdominat at the start of inflation, it will remain so for the entire inflationary period, so that the vector field does not affect the inflationary dynamics.

We assume that the scaling of $f$ and $m$ during inflation is terminated at the end of inflation i.e.
\begin{equation}
f=1,\quad\textrm{and}\quad M=\textrm{constant},
\end{equation}
so that $\alpha=0$. Therefore, the energy density can scale in two ways after inflation (as shown in Ref. \cite{27}):
\begin{enumerate}
\item If the field is light at the end of inflation ($M\ll H$), the energy density scales as $\rho_{W}\propto a^{-4}$; therefore, the energy density scales as relativistic particles (a radiation fluid). This shows a remarkable contrast with the scalar field case where the energy density remains constant even after inflation.
\item If the field is heavy at the end of inflation ($M\gg H$), the energy density scales as $\rho_{W}\propto a^{-3}$; therefore, the energy density scales as non-relativistic particles (a matter fluid) where the average pressure is zero. The oscillating vector field behaves as an isotropic material and can dominate the Universe without generating too much statistical anisotropy. This is an important feature in the vector curvaton scenario because it can produce the curvature perturbation and the vector field can dominate (or nearly dominate) the Universe without inducing excessive anisotropic expansion.
\end{enumerate}

\section{STATISTICAL ANISOTROPY}
An alternative to parameterize the presence of statistical anisotropy corresponds to a general change in the definition of the spectrum of $\zeta$ as shown in Eq. (\ref{eq2}), where the persistence of statistical anisotropy in the observational data strongly suggests the presence of a vector field in the inflationary epoch. Ref. \cite{27} shows that the statistical anisotropy can be parameterized as
\begin{eqnarray}\label{3.107}
g_{_{\zeta}}=\xi\left( \frac{{ P}_{\parallel}-{ P}_{+}}{{ P}_{\phi}+\xi{ P}_{+}}\right)=\left( \frac{\xi}{1+\xi}\right)\left[ \left(\frac{3H_{*}}{{M}}\right)^{2}-1 \right],
\end{eqnarray}
where $\xi\equiv\left(N_{W}/N_{\phi}\right)^{2}$, $N_{W}\equiv \partial N/\partial W$ and $N_{\phi}\equiv \partial N/\partial\phi$ are the derivatives of the amount of expansion with respect to the vector field and scalar field respectively, and $P_{\phi}$ is the spectrum of the scalar field present during inflation. These quantities are defined within the $\delta N$ formalism \cite{18}.

On one hand, we want the vector field to be in charge of generating the curvature perturbation in $\zeta$ to dispense with the scalar fields. In the case that the vector field dominates after inflation, $N_{W}\gg 1$ and $\xi\gg 1$ and, therefore,
\begin{eqnarray}\label{3.108}
g_{\zeta}\approx\left(\frac{3H_{*}}{{M}} \right) ^{2}\gg 1,
\end{eqnarray}
which is in disagreement with the observations. On the other hand, if the contribution of the vector field perturbations to $\zeta$ is subdominant compared to the scalar field, then $\xi\ll 1$, which implies losing the idea of the curvaton field as the generator of the curvature perturbation.

In the case $f\propto a^{-4}$, the mass of the vector field can be increased sufficiently so that $M\sim H$; thus, the spectra of the longitudinal and transverse modes are almost of the same order:
\begin{eqnarray}
\left. 
\begin{array}{cc}
{ P}_{\parallel}=\left(\frac{3H_{*}}{{M}} \right) ^{2}\left( \frac{H_{*}}{2\pi}\right)^{2}&\\
&\\
{ P}_{+}=\left(\frac{H_{*}}{2\pi}\right) ^{2}&
\end{array}
\right\rbrace\quad\Rightarrow\quad { P}_{\parallel}\sim{ P}_{+}={ P}_{\phi},
\end{eqnarray}
so that the curvature perturbation is almost isotropic, allowing us to dispense with the scalar field and to assume that the curvature perturbation is dominated by the contribution of the vector field; therefore, the expression for the statistical anisotropy in Eq. (\ref{3.107}) can be written as
\begin{eqnarray}\label{3.111}
g_{\zeta}\approx\frac{{ P}_{\parallel}}{{ P}_{+}}-1=\frac{\delta{ P}}{{ P}_{+}},
\end{eqnarray}
where  $\delta{ P}/P_+=({ P}_{\parallel}-{ P}_{+})/P_+$  is the fractional differential spectra. If this fractional difference is not excessive, the vector field could generate statistical anisotropy in the CMB within the observational bounds.
\section{Conclusions}
We have studied a massive vector field present during inflation, inspired by the scalar curvaton scenario, which involves a non-canonical kinetic term in the Lagragian, where the mass and kinetic functions vary as functions of time. The parameterization of the kinetic function is of the form $f\propto a^{-1\pm3}$, where $a(t)$ is the expansion parameter. In the case of $f\propto a^{2}$, the mass of the vector field is constant; in this regime the vector field will be subdominant during inflation and also its contribution to the primordial curvature perturbation $\zeta$; in this case, the vector field has the ability to generate substantial statistical anisotropy making the role for which the vector curvaton is invoked be lost. In contrast, when $f\propto a^{-4}$ is chosen, the vector field mass increases with time; thus, at the end of inflation, the vector field can be heavy and can dominate the energy density on the Universe generating this way $\zeta$; in addition, it can generate the statistical anisotropy that would be within the observational bounds if the fractional differential spectra is not excessive.


\begin{theacknowledgments}
We would like to thank the organizers of the IX Mexican School of the DGFM-SMF for his hospitality and attention.
A. A. N. acknowledges support for mobility from VIE (UIS) grant number 2012010624. Y. R. acknowledges support for mobility from VCTI (UAN). A. A. N. and Y. R. are supported by Fundaci\'on para la Promoci\'on de
la Investigaci\'on y la Tecnolog\'ia del Banco de la Rep\'ublica
(COLOMBIA) grant number 3025 CT-2012-02. In addition, Y. R. is
supported by VCTI (UAN) grant number 2011254. This work is entirely based on Refs. \cite{81,27,28,29} and is the result of A. A. N.'s MSc thesis.
\end{theacknowledgments}



\bibliographystyle{aipproc}   




\end{document}